\documentclass[%
aip, preprint,
amsmath,amssymb]{revtex4-2}

\usepackage[utf8]{inputenc}
\usepackage[T1]{fontenc}
\usepackage[english]{babel}

\usepackage{graphicx}
\usepackage{amssymb}
\usepackage{amsmath}
\usepackage{siunitx}
\usepackage{subcaption}
\usepackage{paralist}
\usepackage{nth}
\usepackage{todonotes}

\usepackage{tikz}

\usepackage{cleveref}

\graphicspath{{figures/}}
\begin{document}

\title{Electric field measurements on the INCA discharge}

\author{Christian L\"utke Stetzkamp}
\email{Christian.LuetkeStetzkamp@rub.de}
\author{Tsanko V. Tsankov}
\email{Tsanko.Tsankov@rub.de}
\author{Uwe Czarnetzki}
\affiliation{Institute for Plasma and Atomic Physics, Ruhr University Bochum, D-44780 Bochum, Germany}

\date{\today}

\begin{abstract}
Developing large-area inductively coupled plasma sources requires deviation from the standard coil concepts and the development of advanced antenna designs. First steps in this direction employ periodic array structures. A recent example is the Inductively Coupled Array (INCA) discharge, where use is made of the collisionless electron heating in the electric field of a periodic array of vortex field. Naturally, the efficiency of such discharges depends on the how well the experimental array realizes the theoretically prescribed field design.

In this work two diagnostic methods are employed to measure the field structure of the INCA discharge. \textit{Ex situ} B-dot probe measurements are compared to \textit{in situ} radio-frequency modulation spectroscopy (RFMOS) and good agreement between their results is observed. Both diagnostics show systematic deviations of the experimentally generated field structure from the one employed in the theoretical description of the INCA discharge. The subtleties in applying both diagnostic methods together with an analysis of the possible consequences of the non-ideal electric field configuration and ways to improve it are discussed. 
\end{abstract}

\maketitle

\section{Introduction}

Operation of large area inductively coupled (ICP) discharges at low pressures
($p \le \SI{100}{\pascal}$) is very important for various
applications~\cite{Lieberman2005,Hopwood1992, Okumura2010}, because these
discharges can provide large ion fluxes needed for anisotropic (directional)
surface treatment. However, standard ICP discharges are not ideally suited for
providing a homogeneous plasma over large areas ($\sim\SI{1}{\square\meter}$) at
low pressures. An approach to achieve this is to choose an antenna configuration
different to the standard flat spiral coil. Several other designs have been
developed over the years. The most common alternatives are probably solenoid or
dome-shaped antennas~\cite{Hopwood1992, Okumura2010, Godyak2011}. Also some
concepts using multiple antennas have been proposed. One of the more recent
developments is to replace the coils by linear
rods~\cite{Guittienne2017,Hollenstein2013,Lecoultre2012}. Apparently, this
allows for easy up scaling. However, rather high currents are required to
compensate for the comparably low electric field produced by a linear rod.

Recently another possibility was investigated. It is based on a theoretical
concept for collisionless resonant electron heating proposed
by~\citet{Czarnetzki2014}. Under collisionless conditions (low pressure) an
infinite array of electric RF vortex fields each a distance $L$ apart and
oscillating at the frequency $\omega_0$ can couple energy efficiently to the
electrons. A two-dimensional array of small planar spiral coils realizes such a
structure. When the RF current through the coils has the same phase, the
structure of the field repeats itself after a distance $\Lambda = L $. The
electrons moving at a velocity of $v_{\text{p}}=\Lambda\omega_0/2\pi$ parallel
to the plane of the array see a constant electric field and are accelerated
constantly. Because of the vortex structure of those fields, these electrons
never run out of resonance.

This concept was experimentally realized as the INductively Coupled Array (INCA)
discharge. It shows consistently a reliable and efficient operation at pressures
as low as \SI{0.1}{\pascal}~\cite{Ahr2018,Ahr2018a}. This work was accompanied
by a more in-depth theoretical analysis of this electron heating
mechanism~\cite{Czarnetzki2018}. It revealed, that the efficient operation is
caused by the fact, that all electrons participate in the heating process,
making it more efficient than the stochastic heating in standard ICP discharges.
By including also the electron collisions with the background gas the theory
reveals that for low pressures the collisionless non-local heating prevails and
at higher pressures the standard local Ohmic heating
dominates~\cite{Czarnetzki2018}. At the pressure for which the mean free path of
the electrons $\lambda$ is comparable to the characteristic length of the array $\Lambda$:
$\lambda \sim 0.1\Lambda$, the transition between the two regimes
occurs~\cite{Czarnetzki2018}. Additionally to the electric field configuration
already described, called the ortho-array, another field configuration was
proposed and discussed in the theoretical works on the INCA discharge -- the so
called para-array, where the phase of the current changes by \SI{180}{\degree}
between the neighboring coils~\cite{Czarnetzki2014}. Recently, also this field
structure was realized experimentally and it was shown its behavior is very
similar to that of the ortho-array~\cite{Luetkestetzkamp21}. The differences are in good agreement to
the theoretical predictions.

In the INCA discharge, the exact local structure of the electric field is
vital for its operation in the stochastic mode. In the theoretical works, ideal circular coils where assumed, which produce a purely vortex electric field. However, the practical realization of such coils is not possible, especially for for small coils where the in- and out-connections already lead to a notable deviation of the induced field.
Additionally the coils used in the experiments are spiral ones, because they
have two windings to enhance the electric field. Till now the structure of the
electric field produced by the coils used in the experiment where unknown. In
this work, the structure of the electric field is obtained using two different
diagnostics. In-situ measurements with RF-modulation spectroscopy (RFMOS) are
compared to B-dot measurements (without plasma). The RFMOS measurements also
allow an insight into the spatially resolved strength of the residual capacitive
coupling in the discharge.

The paper is organized as follows: first the theoretical background of both
diagnostics is described and the experimental setup is shown.
In section \ref{sec:dataeval} the numerical evaluation scheme for the B-dot
measurements is outlined. Then the experimental results are shown and both
diagnostics are compared in section \ref{sec:results}. The conclusions with
final remarks and an outlook follow in section \ref{sec:conclusion}.

\section{Theory}
\label{sec:theory}

\subsection{RF modulation spectroscopy (RFMOS)}

To obtain \textit{in situ} information about the amplitude and the phase of the induced
electric field as well as information about the strength of the capacitive
coupling compared to the inductive coupling into the plasma an optical
diagnostic called radio frequency modulation spectroscopy (short: RFMOS) is
used~\cite{Crintea2008,Celik2011,Tsankov2011}. For this diagnostic the temporal
modulation of the emission from a chosen line (from now on called modulation) is
measured using an ICCD camera.

For the further analysis a derived quantity is used, the normalized time
dependent modulation $\eta(t)$ during one RF period. It is defined by the ratio
of the emission intensity $I(t)$ to its value averaged over one RF period:
\begin{align}
  \label{eq:eta}
  \eta(t) = \frac{I(t)}{\langle  I(t) \rangle} -1.
\end{align}
By using the normalized modulation: the influence of many effects on the
measurement is eliminated: camera adjustments (exposure time, gain), background
light and inhomogeneity of the electron and the neutral gas density. With the
RFMOS technique information about the plasma is obtained from the amplitude and
the phase of the different Fourier components of $\eta(t)$. However, the DC
component of $\eta(t)$ is always zero due to its definition.

The temporal modulation of the emission intensity arises due to the following
reason. In a radio frequency discharge, the time dependent electric field
modulates the EEDF, which in turn modulates the excitation to a higher atomic
state with excitation energy $\epsilon_{\text{exc}}$ and lifetime $\tau$. When
the population of an excited state is modulated, consequently also the
spontaneous emission from that state is modulated. This is the fact used by the
RFMOS diagnostic technique. This modulation happens mainly at two frequencies:
At the first harmonic of the applied RF frequency $\omega_{\text{RF}}$, due to
the fact, that the antenna acts as an electrode and there is some capacitive
coupling. And at the second harmonic due to the inductive coupling. Because of
this, the modulation of the emission intensity in an inductively coupled
discharge should be mainly sinusoidal, but can contain some deviations caused by
residual capacitive coupling. To separate the components of the modulation at
the different frequencies, a Fourier analysis of $\eta(t)$ is performed:
$\mathcal{F}_{n\omega}[\eta(t)] = \hat\eta_{n\omega}$, where $n$ gives the
number of the harmonic of the fundamental frequency. Here this is the frequency
of the applied RF signal $\omega_{\text{RF}}$.

To obtain a relation between the intensity modulation $\hat \eta_{n\omega}$ and
the electric field the following approach is applied. The rate equation for the
excited state is combined with the assumption of a time independent, isotropic
near Maxwellian distribution of the electrons that is displaced by a small, time
dependent oscillation velocity~\cite{Crintea2008,Celik2011,Tsankov2011}. From
this, the measured modulation $\hat\eta_{(1,2)\omega}$ can be expressed as
follows:
\begin{align}
  \label{eq:eta1}
  |\hat\eta_{1\omega}| &=
                         \frac{1}{\sqrt{1+{(\omega\tau)}^2}}\left(\frac{4}{3}
                         \frac{\epsilon_{\text{exc}}}{k_{B}T_e}-2\right)
                         \frac{\vec u_d\cdot\vec{u}_{\text{osc}}}{v_{\text{th}}^2} \\
  \label{eq:eta2}
  |\hat\eta_{2\omega}| &= \frac{1}{\sqrt{1+{(2\omega\tau)}^2}} \left(
                         \frac{1}{3}
                         \frac{\epsilon_{\text{exc}}}{k_{B}T_e} -
                         \frac{1}{2}\right) \frac{\vec u_{\text{osc}}^2}{v_{\text{th}}^2}
\end{align}
where $v_{\text{th}}$ is the thermal velocity of the electrons and
$\vec u_d$ their DC drift velocity.

Under locality conditions the oscillation velocity $u_{\text{osc}}$ is
determined by the local induced electric field:
\begin{align}
  u_{\text{osc}} &= \frac{eE}{m_e\sqrt{\omega_{\text{RF}}^2 + \nu_{\text{m}}^2}}
\end{align}
with $\nu_{\text{m}}$ the electron collision frequency. Due to this relation,
the amplitude of the second harmonic of the modulation ($\hat\eta_{2\omega}$) is
proportional to the square of the induced electric field.

Usually the amplitude of the modulation is weak (a few percent). In an
inductively coupled discharge, it is even smaller compared to
capacitive discharges~\cite{Kadetov2004}. To measure the small change
in the plasma emission, the detector must be very sensitive and the
spectral line used should be chosen carefully. The upper state of the
observed transition should not be populated from metastable states or
by cascades, i.e.\ it has to be populated predominantly from the ground state. This
requirement is specially important, because the excitation from the
ground state to a high state is mainly determined by the high
energetic tail of the EEDF\@. This part of the distribution function
also is the one that is most affected by the variation of the induced
electric field. Due to this, the modulation of the emission from such
a highly energetic state is stronger compared to the one from a low
energy state. This can also be seen in the equations
(\cref{eq:eta1,eq:eta2}), the amplitudes of the first and second
harmonic increases with the excitation energy
$\epsilon_{\text{exc}}$. The other requirement is, that the lifetime
of the excited state should be much shorter than the RF period
($\tau_{\text{rf}}=\SI{74}{\ns}$), so that no averaging over time
happens. This is also seen in~\cref{eq:eta1,eq:eta2}, where  the amplitude
decreases for increasing values of $\tau$. For the measurements in hydrogen
performed in this work, the H$_\alpha$ line is used~\cite{Tsankov2011}.

\subsection{B-dot measurement}

To determine the electrical field of a coil or an arrangement of coils first the
magnetic field is measured. This is done by inserting a conductive loop oriented
towards $\hat e_i$ into the field and measuring the induced voltage. In the
following only fields produced by flat coils in the x-y-plane are considered.

The voltage $U_{\text{ind}}$ induced by a time dependent magnetic flux $\phi$
into the conductive loop is given by Faraday's law:
\begin{align}
  U_{\text{ind}} &= -\frac{\mathrm{d}\phi}{\mathrm{d}t} = - \frac{\mathrm{d}}{\mathrm{d}t}\int \vec B\cdot \mathrm{d} \vec{S} 
\end{align}
Here $\vec B$ is the magnetic field, $\vec S = S\vec n$ with $S$ the area of
the conductive loop and $\vec n$ the normal vector of its area. If the conductive
loop is oriented towards $\hat e_i$ and has $N$ windings:
\begin{align}
  U_{\text{ind}} &= -N \frac{\mathrm{d}}{\mathrm{d}t} B_i S
\end{align}
where $B_i$ is the $i$ component of the mean magnetic field inside the
conductive loop. When the loop is stationary and the magnetic field is harmonic
in time ($B\propto e^{i\omega t}$) it follows
\begin{align}
  \label{eq:B}%
  B_i &= - \frac{U_\text{ind}}{i\omega N S}
\end{align}
To measure the full magnetic field this measurement must be preformed separately
three times, once with the loop oriented towards each coordinate axis.

To calculate the electric field $\vec E$ from the measured magnetic field, there
are in principle two options: Using the curl of the magnetic field or using the
rotation of the electric field. While the first one seems to be the superior one, because it
assumes no knowledge about the form of the electric field, it has some numerical
difficulties. Assuming also the electrical field to be harmonic, from the curl
of $\vec B$ follows: $\vec E = c^2/(i\omega) \nabla\times\vec B$. In the curl
there small differences of rather large quantities are calculated and that is
numerically unstable. Because of this, the curl of the electric field Ansatz is
promising. However, here one assumption is needed: $E_z = 0$, this approximation
if fine, because the coils are in the x-y-plane and thereby there are no
currents in z-directions, which could cause electric fields in z-direction.
Assuming that also $E_x,E_y$ vanish at $z\to \infty$ it follows
\begin{align}
  \label{eq:Ex}
  E_x &= -i\omega \int_\infty^z B_y \mathrm{d}z \\
  \label{eq:Ey}
  E_y &= i\omega \int_\infty^z B_x \mathrm{d}z
\end{align}

The consistency of the measured magnetic field would normally be checked by
using $\nabla\cdot \vec B \overset{!}{=} 0$. But in this calculation again small
differences of large quantities appear, so it is numerically unstable.
Nevertheless, every curl-field is divergence free, so it is sufficient to test
for $\nabla\times\vec E = -i\omega \vec B$. The $x$- and $y$-component of this
equation were already used to calculate the electric field, so that it is
sufficient to check the $z$-component:
\begin{align}
  \label{eq:Bself}
  \partial_xE_y - \partial_y E_x &\overset{!}{=} -i\omega B_z
\end{align}
Here $B_z$ is calculated from $E_{x,y}$ and than compared to the independently
measured $B_z$ field. If both are the same, the divergence of the magnetic field
vanishes and the diagnostic is consistent. However, this does not test the
assumption of $E_z=0$, because the contributions of $_z$ to the derivatives of
$E_{x,y}$ cancel out.

\section{Experimental Setup}

\subsection{Discharge Setup}

A more detailed description of the discharge setup can be found
elsewhere~\cite{Ahr2018,Ahr2018a}. Here only a brief description is give. The
INCA discharge consists of a rectangular stainless steel vacuum chamber
(\SI{40x40x25.6}{\centi\meter}). The antenna array is
positioned roughly in the middle of the short side, dividing the volume into two
separate regions. On one side (depth \SI{13}{\cm}) the plasma is created by
supplying a \SI{13.56}{\mega\hertz} RF signal through an L-type matching box to
the antenna. After the matching box, but before the vacuum feedthrough the
current is measured. The power is produced by a Dressler Cesar generator. The
other side houses the wiring and the cooling of the antenna array. The placement
of the array inside the vacuum chamber is to avoid mechanical stress due to
pressure difference over its large area (\SI{40x40}{\cm}).

The antenna array consists of \num{6x6} flat spiral coils, with two windings
each and a thickness of \SI{3}{\milli\meter}. The coils are made of copper and
occupy an area of \SI{5x5}{\centi\meter}, i.e.\ $L=\SI{5}{\centi\meter}$. The
individual coils are connected with each other through copper stripes. The
previous studies~\cite{Ahr2018, Ahr2018a} have shown that for a matching to be
possible, the coil wiring has to be separated into three parallel branches, each
containing \num{12} coils connected in series.

\begin{figure}[htb]
  \centering \includegraphics[width=\columnwidth]{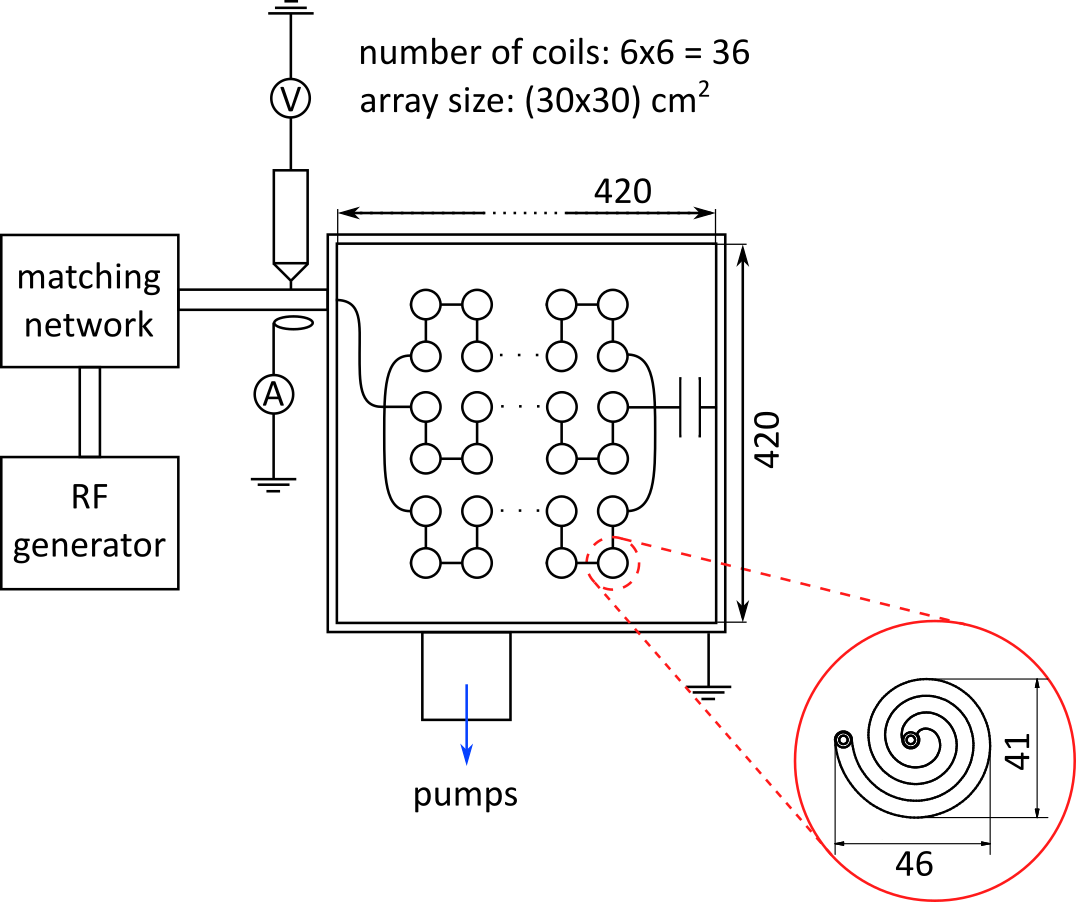}
  \caption{Schematics of the discharge chamber, the dimensions are given in
    mm.}%
  \label{fig:schem:inca}
\end{figure}

The temporally and spatially resolved emission from the plasma is measured using
an ICCD camera (HR16-PicoStar, LaVision) which is placed in front of the antenna
facing lid of the discharge chamber. The antenna facing lid has three windows
aligned on the diagonal of it. In this work mainly the central window and seldom
the upper right window are used.

\subsection{B-dot probe system setup}

\begin{figure}[htb]
  \centering
  \includegraphics[width=\columnwidth]{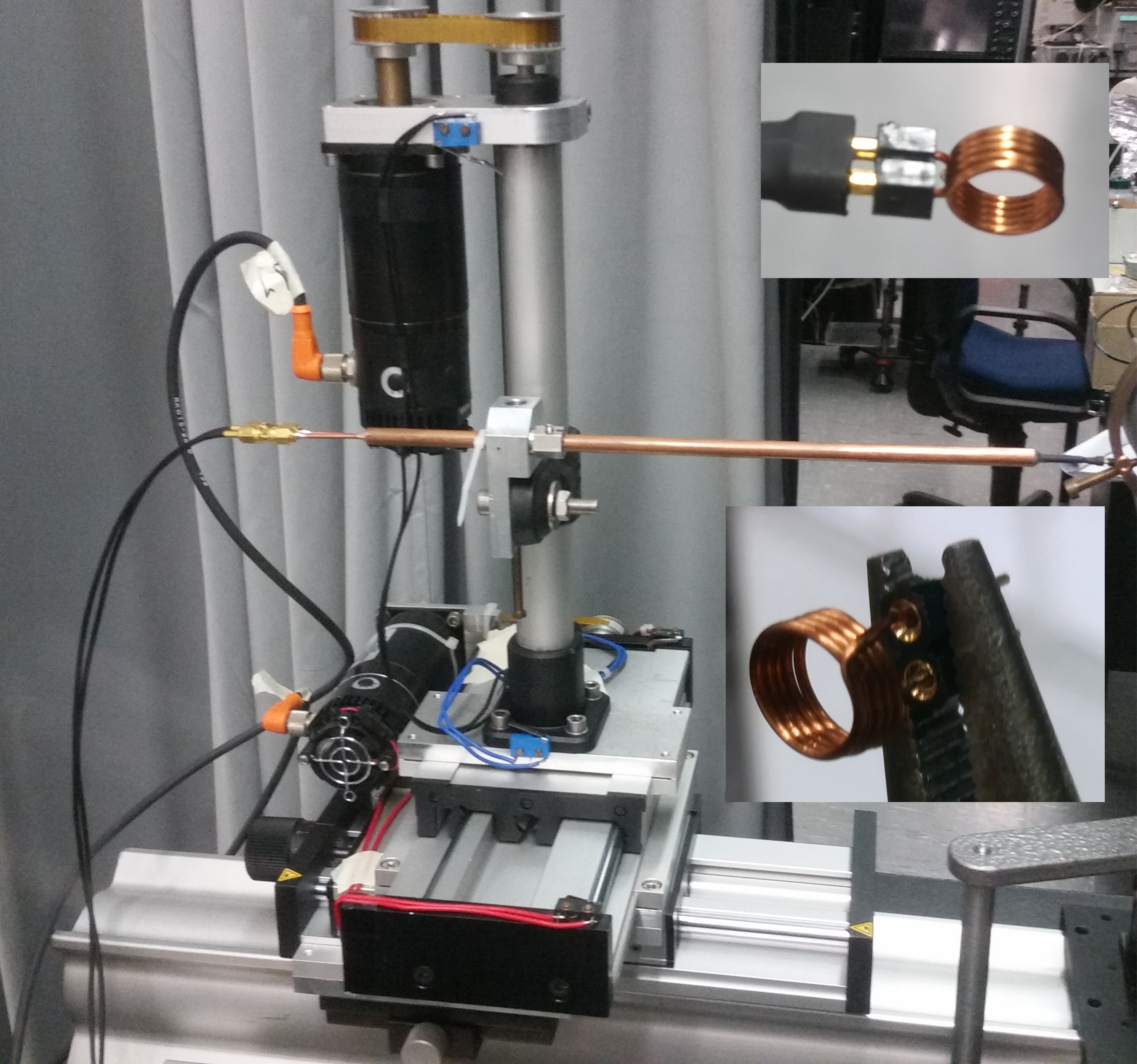}%
  \caption{Photograph of the homemade B-dot probe system with enlarged inlets
    of the pick-up coils for the $x$- and $y$- component of the $B$-field (top) and for the $z$-component (bottom).}%
  \label{fig:b-dot}
\end{figure}

The homemade B-dot probe consists of a copper rod containing two
wires~\cref{fig:b-dot}. At the end of the rod a connector for a pick-up coil is
installed. For the measurement of the $x$- and the $y$-component of the magnetic
field, the same pick-up coil is used, only the rod is rotated. For the
$z$-component a similar coil with different orientation is used. The pick-up
coils are made of copper wire, each have a diameter of \SI{5}{\mm} and \num{5}
windings. To measure the magnetic field not only in one point, but on a grid,
the probe is mounded on a three-axis movement table. Its movement in the
$xy$- plane is controlled by two stepper motors and its movement along the
$z$-axis is controlled manually. The maximum measurement area is \SI{90}{\mm} in
$x$-, \SI{200}{\mm} in $y$- and \SI{100}{\mm} in $z$-direction. 

The ends of the pick-up coil are connected to an oscilloscope (LeCroy wavepro
7100A) using separate coaxial cables. For triggering and comparison also the
current measurement of the chamber is connected to the oscilloscope. To
determine the induced voltage in the pick-up coil, the difference of the
voltages at both ends of the coil is determined on the oscilloscope. High-frequency noise is suppressed by using an internal \SI{20}{\mega\hertz} filter on
all channels. Because the induced voltage is a harmonic oscillation with a known
frequency only its amplitude and its phase are recorded (see following
section). Fluctuations in the generator power are accounted for by recording
also the amplitude of the current through the antenna.  

The measurement is controlled by two self-written programs running on the
oscilloscope. The main program controls the measurement and the movement of the
stepper motors. For the recording itself, the second program is called that controls
the oscilloscope. It records all quantities averaged over \SI{1}{\s}. For the
movement a 3D scan is implemented, that scans each $x$-, $y$-plane in a
serpentine pattern and instructs the user to set the $z$-position accordingly
after scanning each plane. In this work, a grid size of \SI{5}{\mm} in each
direction was used, because this is also the size of the pick-up coils which
effectively limits the maximum spatial resolution.

\section{Data evaluation}
\label{sec:dataeval}

\begin{figure}[htb]
  \centering
  \includegraphics[width=\columnwidth]{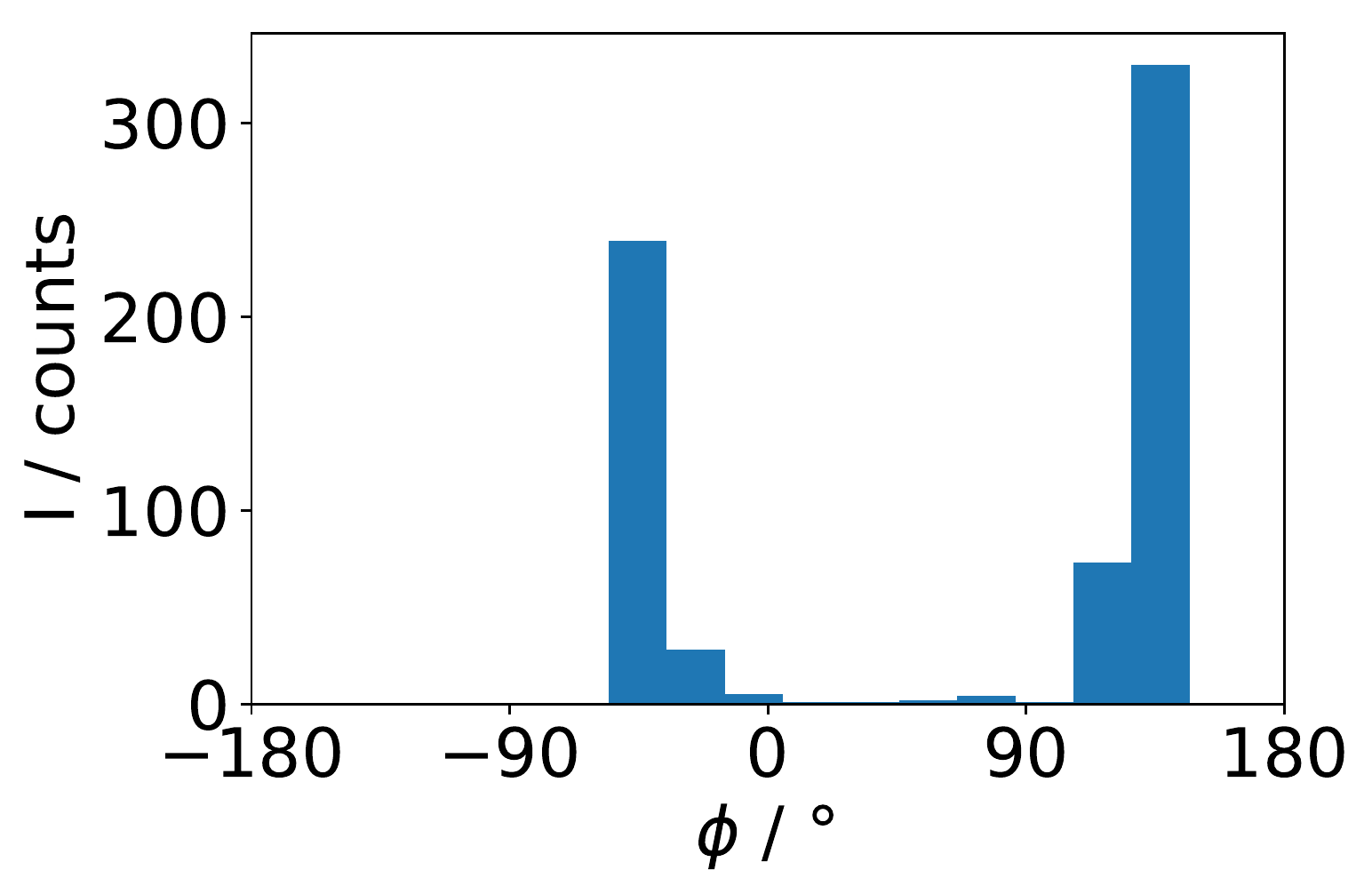}%
  \caption{Example of a histogram of all phases of the induced voltage measured
    in one x-y-plane.}%
  \label{fig:hist}
\end{figure}

In the numerical evaluation of the induced voltage to determine the magnetic
field, also the phase of the induced voltage must be considered. The equation
for a harmonic wave $G$ with a positive amplitude $A(\vec r)$, wave number $\vec k$
and frequency $\omega$ at time $t$ and position $\vec r$ is
\begin{align}
  G(\vec r, t) = A(\vec r)  \exp\left\{i\left(\omega t - \vec k \cdot \vec r +
  \varphi_0 + \varphi_\text{sgn}(\vec r)\right)\right\}
\end{align}
Here $\varphi_0$ is an unknown but constant initial phase and
$\varphi_{\text{sgn}}(\vec r)=\{0,\pi\}$ describes the sign of the field at the position
$\vec r$. The time dependence was not recorded, so that the measured phase is
\begin{align}
  \Delta\varphi(\vec r) &= - \vec k \cdot \vec{r} + \varphi_0 +
                       \varphi_\text{sgn}(\vec r).
\end{align}
The initial phase may be chosen arbitrarily, here we use $\varphi_0=0$. The
magnitude part of the equation can be estimated: $\left|\vec k \cdot \vec
  r\right|= 2\pi r/\lambda$. For a frequency of \SI{13.56}{\mega\hertz} the
wavelength is $\lambda \approx \SI{22}{\m}$ and all distances in the experiment
are smaller than \SI{20}{\cm} and the influence of this part can be neglected.
Because of this, the measured phase contains only the sign of the field. Because
of this, it is necessary to extract the sign from the phase, which is involved,
because the numerical values of the phases contain uncertainties. This is done
for each x-y-plane and each Cartesian component separately. For every plane a
histogram of all phases measured is created (see~\cref{fig:hist} for an example
of such a histogram). In the histogram the highest and the second highest peaks
are found. These peaks should be about \SI{180}{\degree} apart and build the
most likely phases for both signs. Now the phase is shifted, so that the peak at
a higher angle is at \SI{90}{\degree}. By this the peaks should be near
$+\SI{90}{\degree}$ and $-\SI{90}{\degree}$ and the number of measurements where the phase is near
zero should be minimal. The next step is to get the sign of the shifted phases
as a first approximation of the field's sign. However, because of statistical
fluctuations and external distortions, there are still single points, where the
field is weak and the distortions were large enough to determine the wrong sign.
To correct for this use is made of the fact that it is not physical to have single points with a phase that is significantly 
different from that of all surrounding points. Because of this, the sign of all
points, that have a different sign than their x- and y-neighbors, is changed. This
correction is not applied at the edge of the measurement area. Knowing the sign
of the magnetic field components,~\cref{eq:B} is used to determine also the
amplitude of the field components.

The second part of the data evaluation is to calculate the electric field from
the magnetic field. This calculation is done via the curl of the electric field
(see~\cref{eq:Ex,eq:Ey}). For this calculation $B_x$ or $B_y$ must be integrated
with respect to $z$. Numerically this is realized using the Riemann sum
\begin{align}
  \int_\infty^{z_i}G(z')\mathrm{d}z' = \Delta z \sum_{j=i}^N G_j
\end{align}
where $G_j$ is the value of $G$ at $z_j = z_0+j\Delta z$. The obvious advantage
of this method is that the electric field can be calculated for all points where
the magnetic field is known.

To get the calculated z-component of the magnetic field the numerical
derivatives of $E_{x,y}$ are needed. Therefore, the symmetric difference
quotient is used
\begin{align}
  f'(x) \approx \frac{f(x+\Delta x) - f(x-\Delta x)}{2\Delta x}.
\end{align}
Its advantage is that in this way the derivatives are calculated at
the grid points of the function and these grid points are the same for the
measured magnetic field. The disadvantage of it is that the derivatives can not
be calculated at the edge of the measurement area.

\section{Results\label{sec:results}}

\begin{figure*}[htb]
  \centering
  \includegraphics[width=\textwidth]{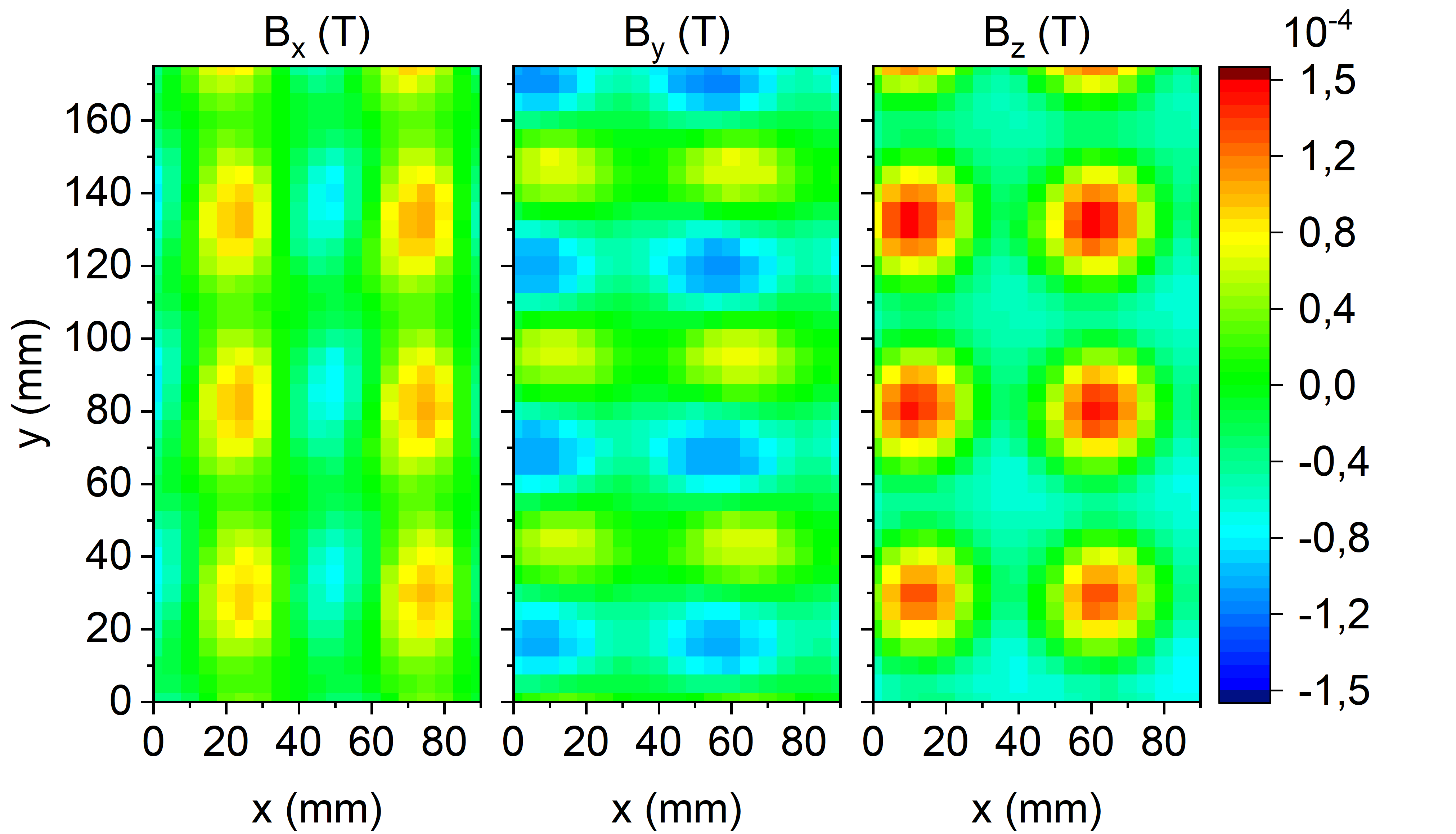}
  \caption{Magnetic field of a part of the INCA antenna array, measured with the
  B-dot probe directly in front of the quartz window.}%
  \label{fig:B0}
\end{figure*}

In~\cref{fig:B0} the magnetic field of a part of the INCA antenna array is
shown. It was measured using the B-dot probe directly in front of the quartz
window. In all three field components the individual coils are clearly visible
and nearly no asymmetry occurs. As one can expect, the $z$-component of the
field is much stronger than the $x$- and $y$-components. This is because the
current through the antenna array is mainly a circulating current and only a
small part of the current is in radial direction. This radial current is caused
by the radius change of the spiral coils. If the coils would be ideal circular
ones, there would be no current component in radial direction.

In the theory section (section \ref{sec:theory}), a procedure was shown to check the
consistency of the B-dot measurements and specially the numerical calculation of
the electric field. The calculated and the measured magnetic field have a very
similar structure. However, the normalized difference shows rather strong
deviations in the proximity of the coil's edges. The reason for those deviations
is apparent: They are at the positions where the spatial dependence of the
electric field components is high. In calculating the $z$-component of the
magnetic field from the electric field components (\cref{eq:Bself}) the
difference of the spatial derivatives of $E_{x,y}$ is calculated. Here, it is a
small difference of large numerical quantities, so the numerical error is high.

\begin{figure*}[htb]
  \centering
  \includegraphics[width=\textwidth]{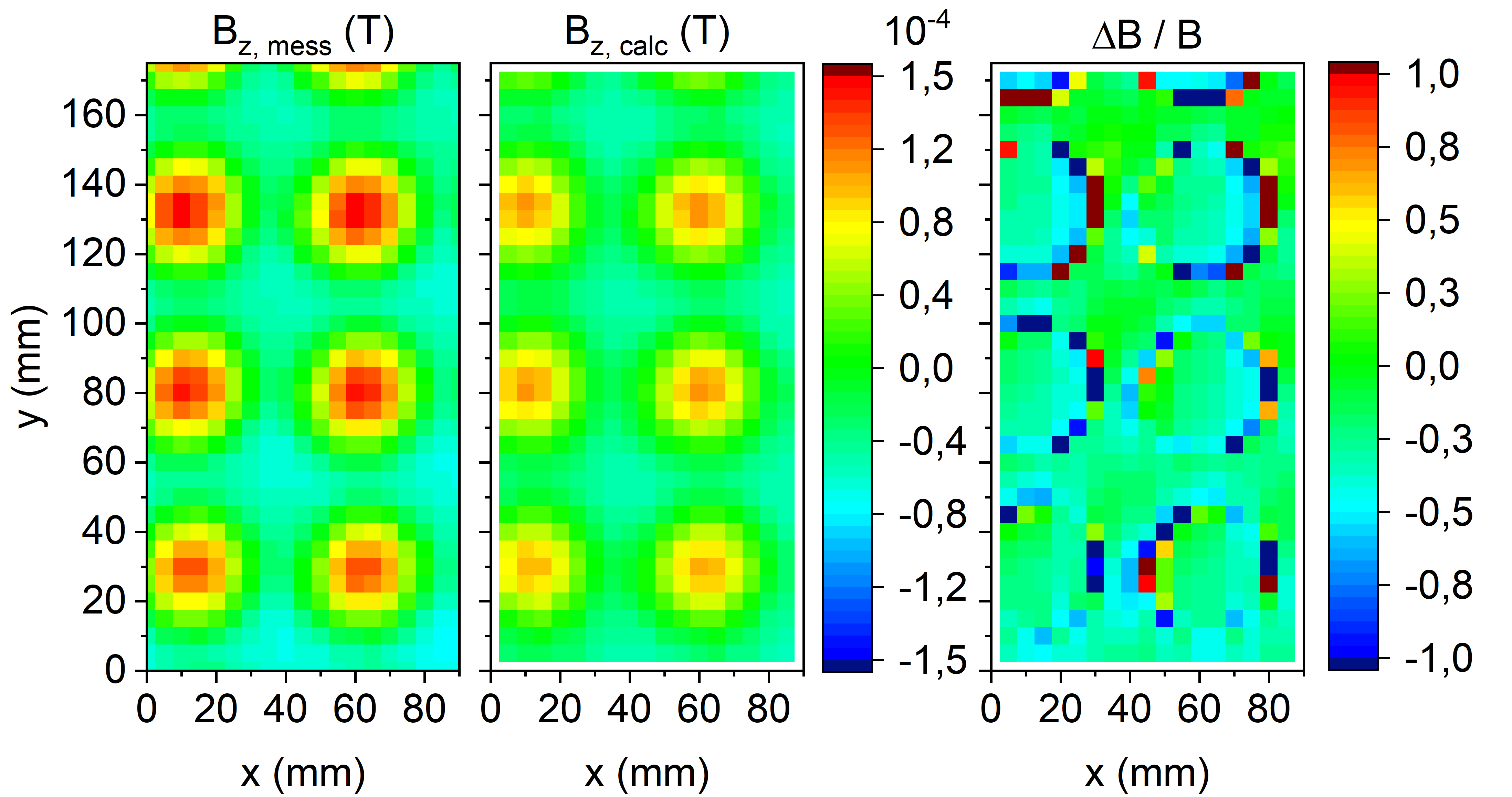}
  \caption{Comparison between directly measured ($B_{z,\text{mess}}$) and from
    the electric field calculated ($B_{z,\text{calc}}$) $z$-component of the
    magnetic field of a part of the INCA antenna array obtained using the B-dot
    probe. Additionally the relative difference is shown.}%
  \label{fig:Bself}
\end{figure*}

The deviations outside of coil edge regions are small and do not show any
structure. There are various reasons, that also there the difference does not
vanish completely. This comparison combines three individual measurements.
Variations of the generator output were compensated by using the current through
the antenna array as a reference, but this might not eliminate all variations.
Also, a different pick-up coil is used for measuring the $z$-component of the
magnetic field. As one can see in~\cref{fig:b-dot} the position of this pick-up
coil relative to the holder is slightly different. Due to this difference in the
measurement girds of the different components of the magnetic field, deviations
between the measured and the calculated $z$-component of the magnetic field
might occur. Because the differences are still small, the consistency of the
diagnostic is confirmed by this comparison.

\begin{figure}[htb]
  \centering
  \includegraphics[width=\columnwidth]{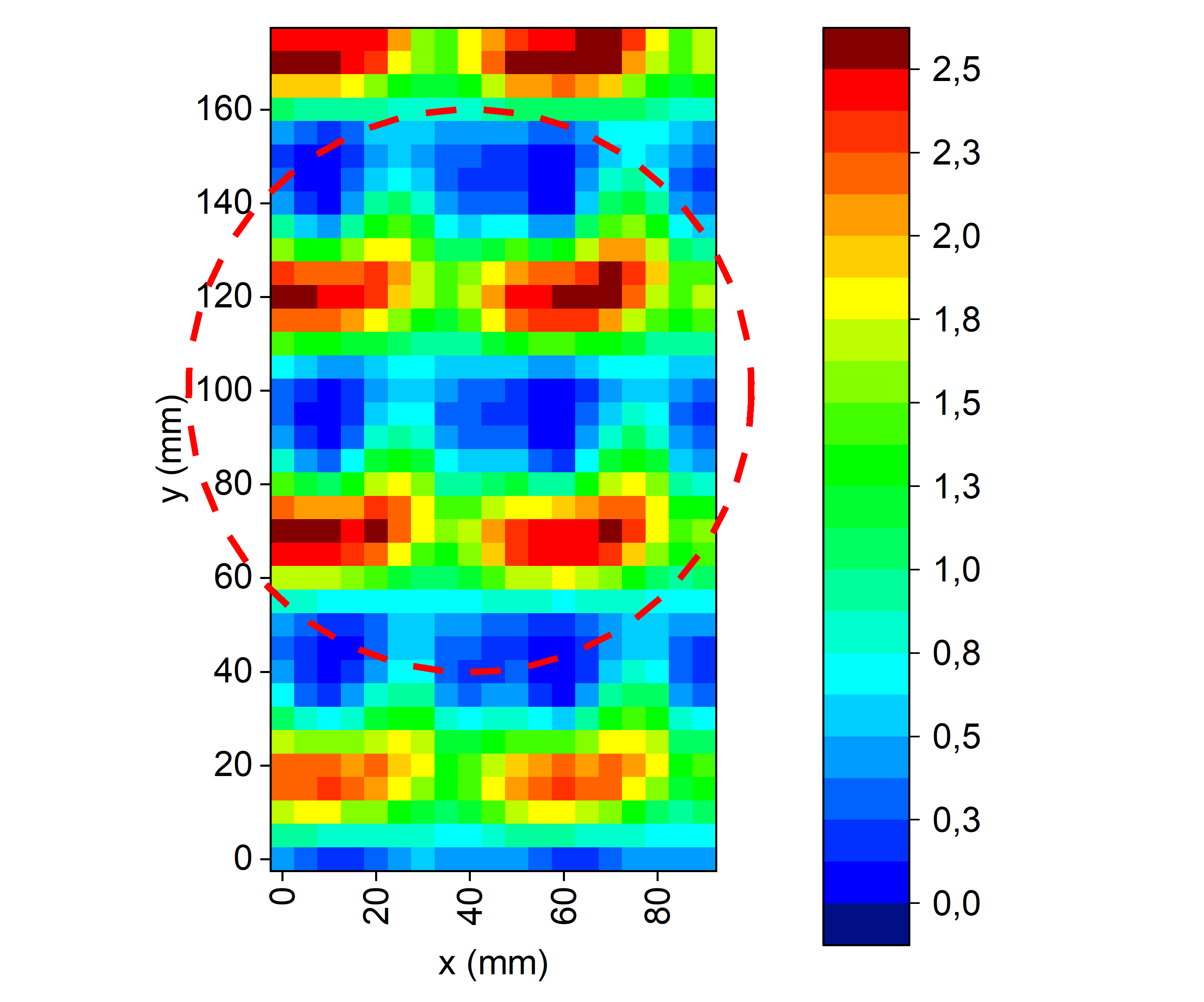}
  \caption{Structure of the square of the induced electric field, $E^2$, from
    the antenna obtained using the B-dot probe. The red circle marks the region,
  that is also accessible to the RFMOS diagnostic at the central window.}%
  \label{fig:Bdot-E}
\end{figure}

The quantity that is more important for the operation of the INCA discharge is
the square of the electric field~\cite{Czarnetzki2018}. Having confirmed, that
the B-dot diagnostic is consistent, it provides \textit{ex situ} (without plasma) access
to the structure of the electric field. The measured square of the electric
field in the direct proximity of the quartz window using the B-dot probe is
shown in~\cref{fig:Bdot-E}. In this figure the red circle marks the region, that
is also accessible to the RFMOS diagnostics at the central window.

The structure of $E^2$ is far from the one created by ideal circular coils,
which was assumed in the theoretical investigations of the INCA
discharge~\cite{Czarnetzki2018,Czarnetzki2014}. Instead the field is strongly
asymmetric and shows kidney-like structures. It is also apparent, that the
different coils influence each other as the intensity of the ``kidneys'' varies
rather strongly between the individual coils.

Using the RFMOS diagnostics it is also possible to measure the square of the
electric field \textit{in situ}. However, it is important to take the locality of the
electrons into account when interpreting the RFMOS measurements. At higher
pressures, where all electrons are local, from the modulation at one point, the
induced electric field at that point can be obtained. At lower pressure, the
electrons become non-local and the modulation at one point is also influenced by
the field at other positions, so that the result smears out spatially.

\begin{figure*}[htb]
  \centering
  \begin{subfigure}[c]{.49\textwidth}
  \includegraphics[width=\columnwidth]{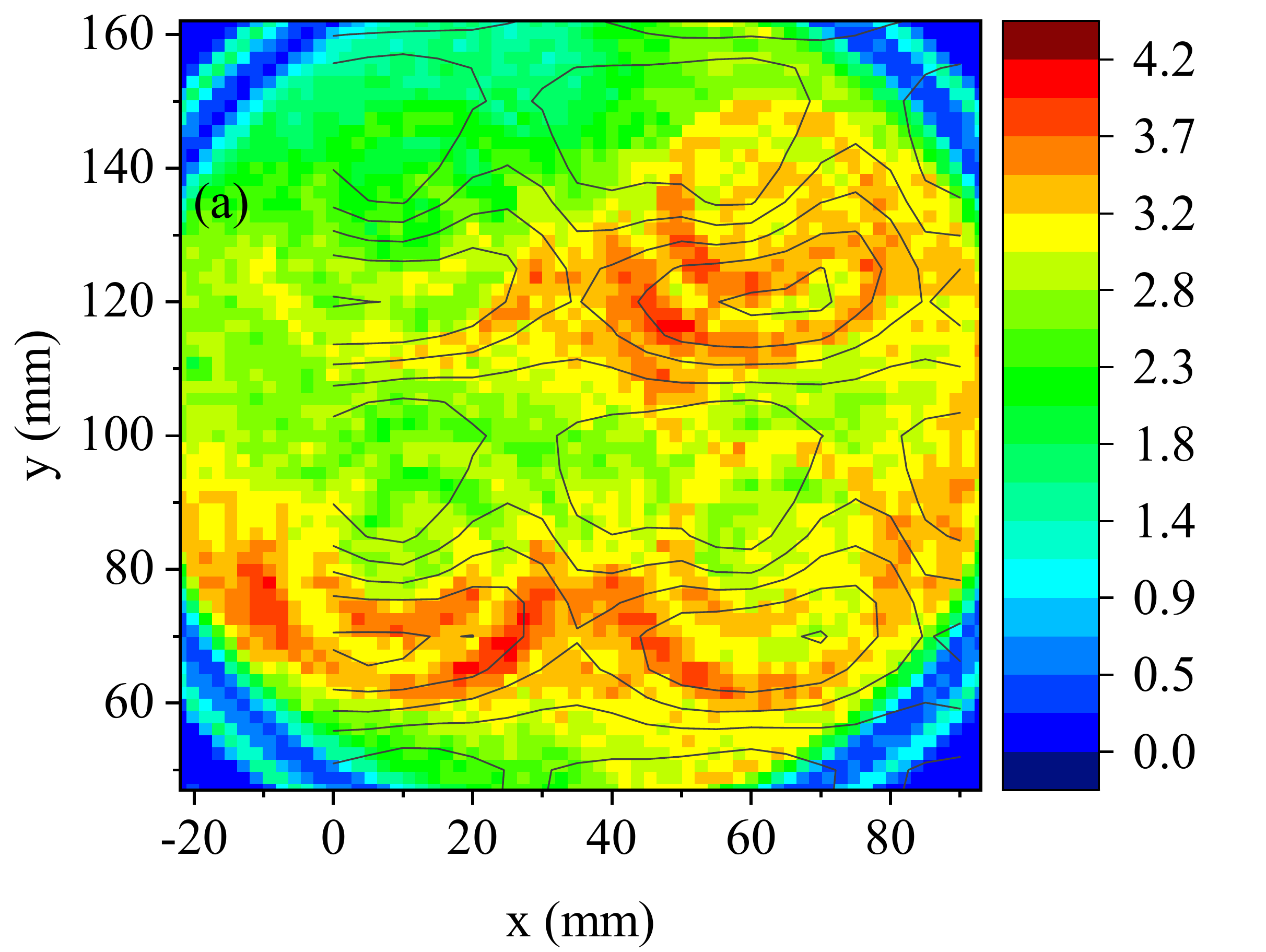}%
  \phantomsubcaption{\label{fig:eta2_amp}}
  \end{subfigure}
  \begin{subfigure}[c]{.49\textwidth}
  \includegraphics[width=\columnwidth]{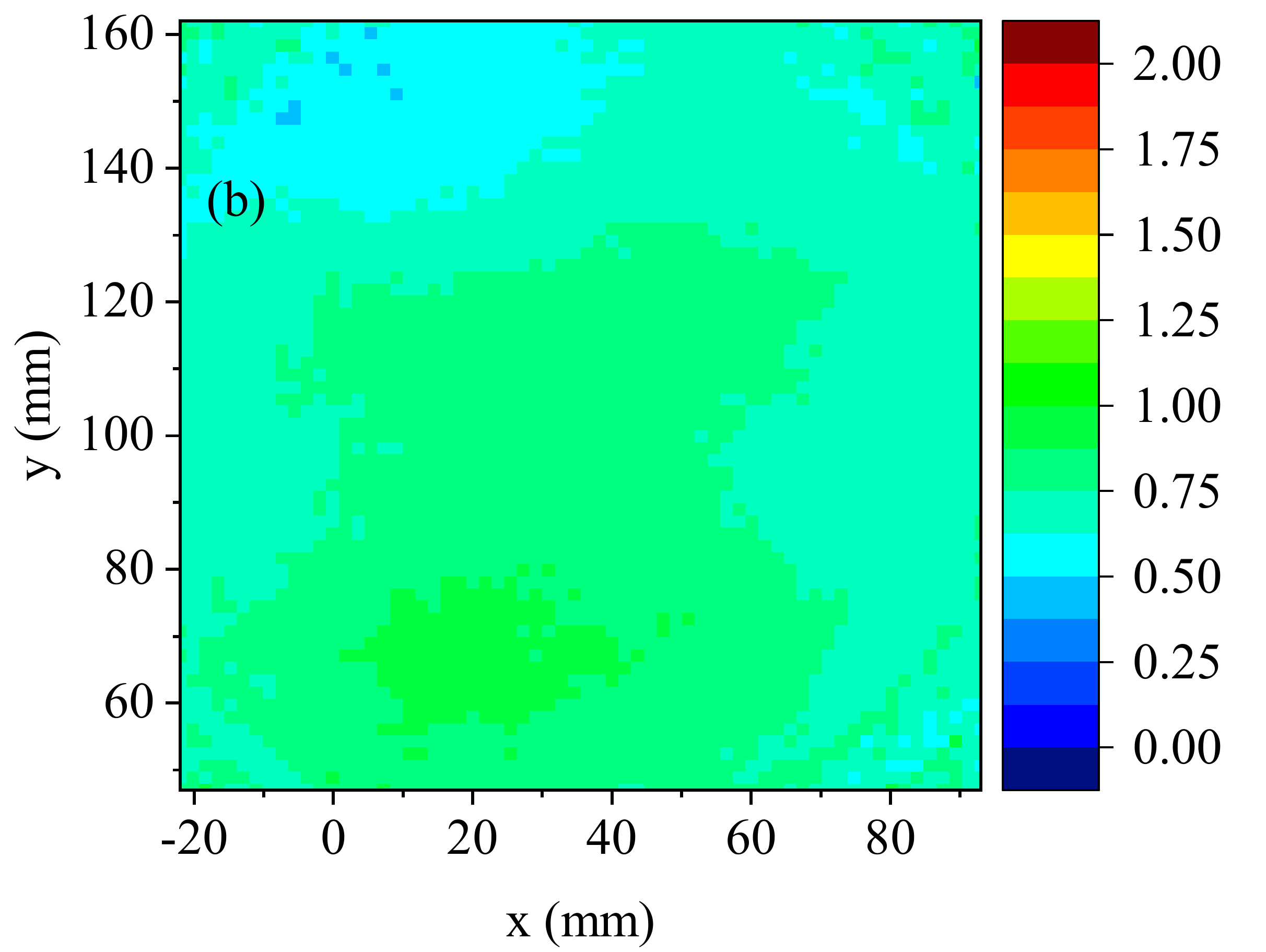}%
  \phantomsubcaption{\label{fig:eta2_phase}}
  \end{subfigure}
  \caption{\subref{fig:eta2_amp} Amplitude and \subref{fig:eta2_phase} phase of
    the second harmonic of the temporal modulation of the H$_\alpha$ line in a
    hydrogen plasma at \SI{15}{\pascal} and \SI{700}{\watt}. The color plots
    show the modulation in percent and the phase in units of $\pi$,
    respectively. The contour lines on the amplitude plot represent $E^2$ from
    the B-dot measurements. The coordinate system is the same as for the B-dot
    measurements (\cref{fig:Bdot-E}).}%
  \label{fig:eta2}
\end{figure*}

To be sure, that the electrons are all local, a measurement is performed at
\SI{15}{\pascal} and \SI{600}{\watt} in hydrogen. At these conditions, the mean
free path of the electrons is few millimeters~\cite{Lieberman2005}. The
modulation of the H$_\alpha$ line was recorded with an ICCD camera through the
middle front window. From this position the central region of $2\times 2$ coils
is visible. The second harmonic of the modulation is proportional to
$u_{\text{osc}}^2$, i.e. the square of the induced electric field,
$E^2$ (see~\cref{eq:eta2}), and is shown together with its phase in~\cref{fig:eta2}.
The figure also shows the values from the B-dot measurements
(\cref{fig:Bdot-E}). The agreement between the two independent diagnostics is
very good. The modulation of the top left coil is relatively weak, but this
could be an optical artifact. The phase of the second harmonic
(see~\cref{fig:eta2_phase}) is relatively homogeneous and shows no structure.
This is not surprising as these four coils are all connected in series.
Therefore, the current and the induced electric field are in phase.

\begin{figure}[htb]
  \centering
  \includegraphics[width=\columnwidth]{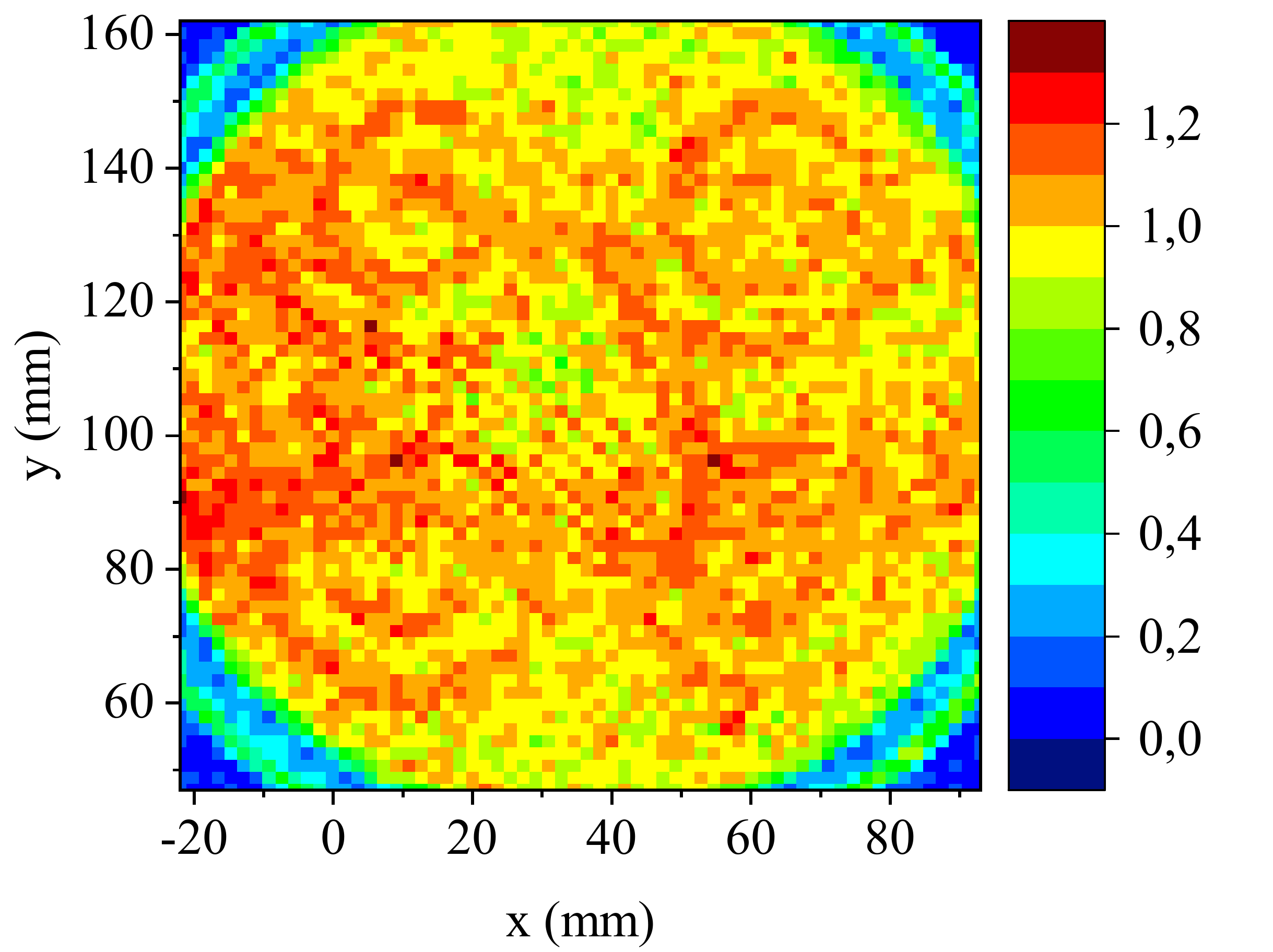}
  \caption{Amplitude of the second harmonic of the temporal modulation of the
    H$_\alpha$ emission in a hydrogen plasma at \SI{0.8}{\pascal} and
    \SI{700}{\watt}. The colors represent the amplitude in percent.}%
  \label{fig:eta2_lowp}
\end{figure}

To confirm that the measurement works as intended and the spatial structures in
the second harmonic of the modulation disappear at lower pressures, a
measurement at \SI{0.8}{\pascal} and \SI{700}{\watt} in hydrogen was performed.
The second harmonic from that measurement is shown in~\cref{fig:eta2_lowp}. As
expected, nearly no structure remains.

\begin{figure*}[htb]
  \centering
  \begin{subfigure}[c]{.49\textwidth}
    \includegraphics[width=\columnwidth]{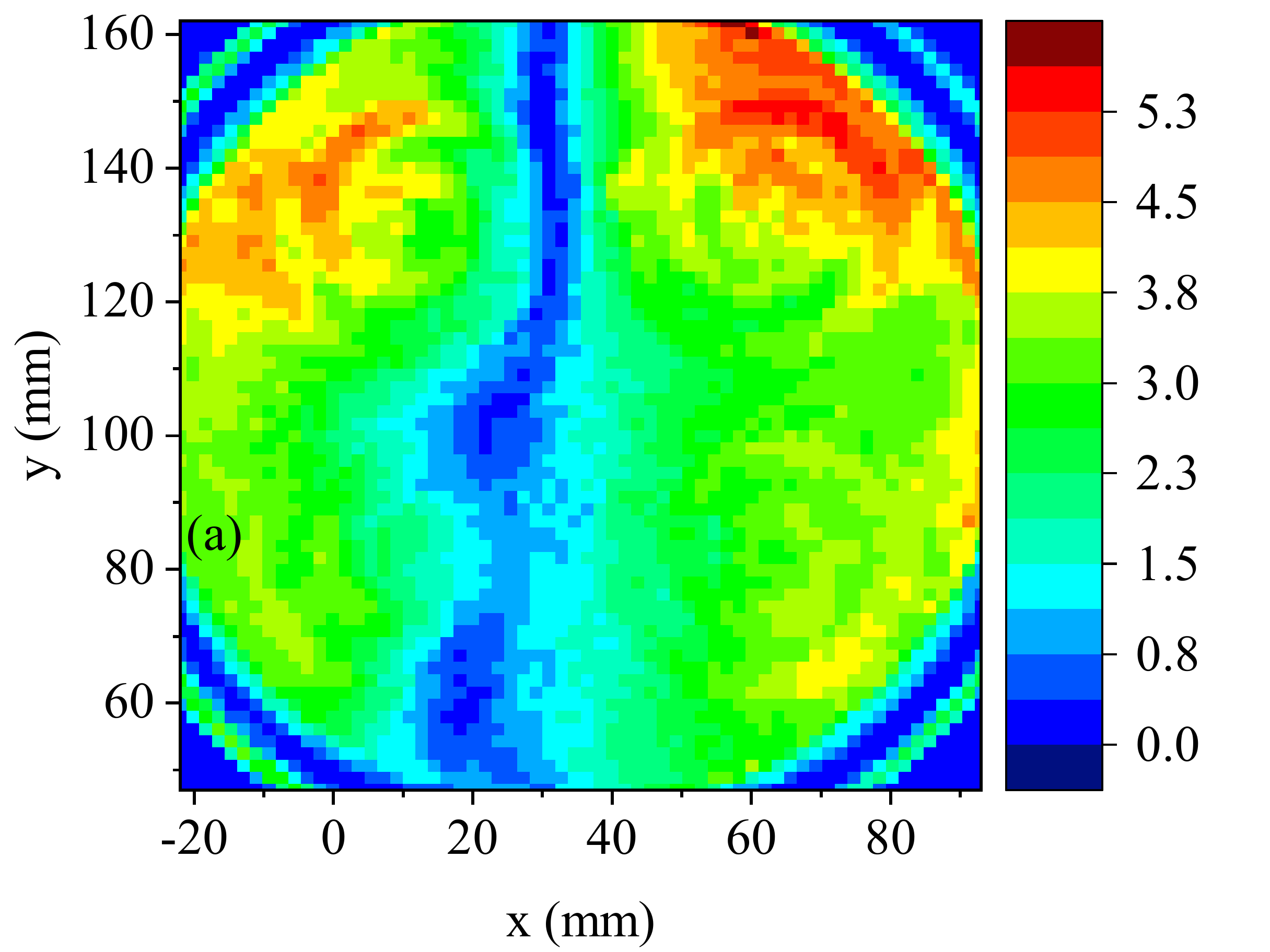}%
    \phantomsubcaption{\label{fig:eta1_amp}}
  \end{subfigure}
  \begin{subfigure}[c]{.49\textwidth}
    \includegraphics[width=\columnwidth]{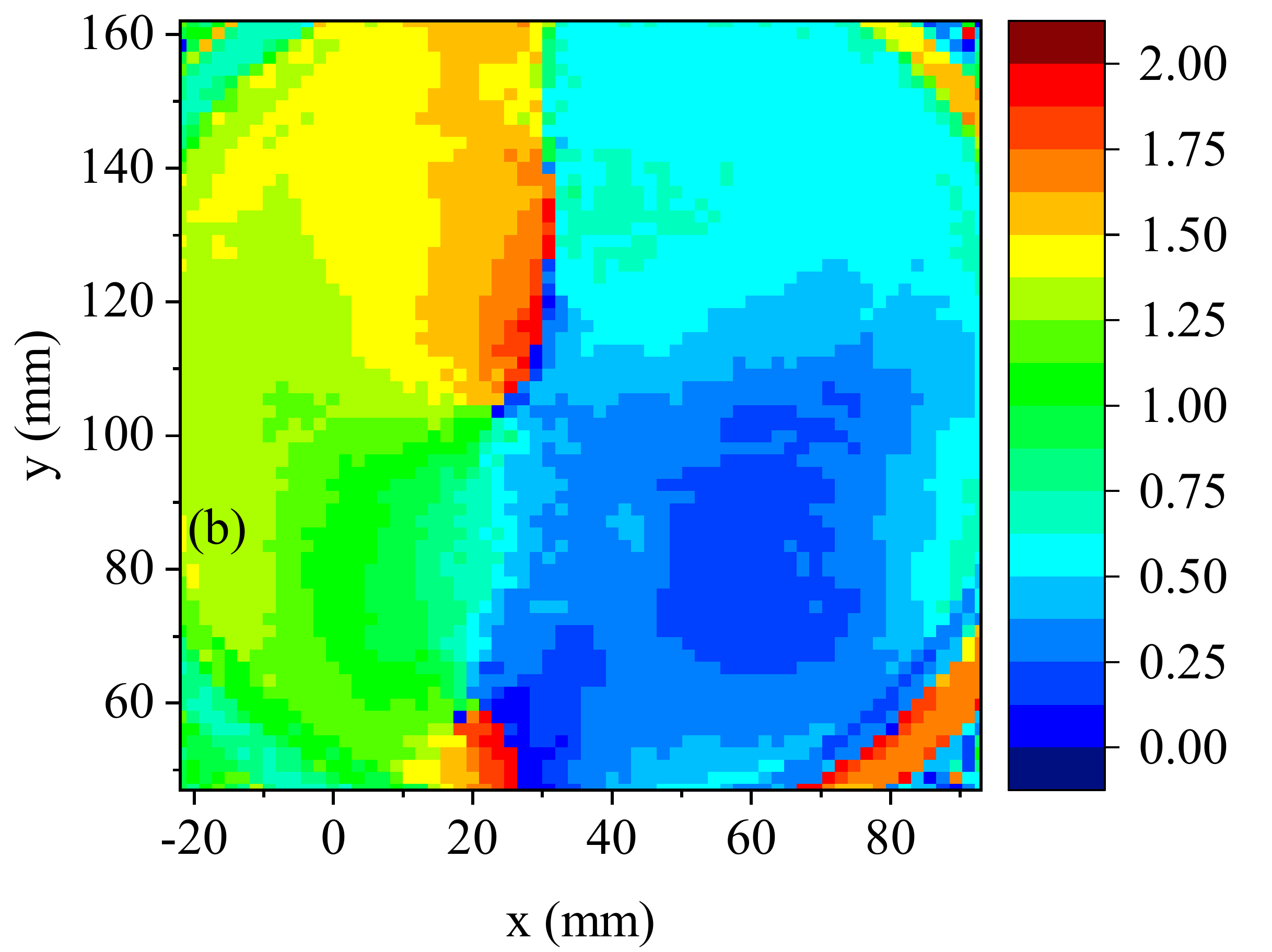}%
    \phantomsubcaption{\label{fig:eta1_phase}}
  \end{subfigure}
  \caption{\subref{fig:eta1_amp} Amplitude and \subref{fig:eta1_phase} phase of
    the first harmonic of the H$_\alpha$ emission in a hydrogen plasma at
    \SI{15}{\pascal} and \SI{700}{\watt}. The color plot represents the
    amplitude in percent and the phase in units of $\pi$, respectively.}%
  \label{fig:eta1}
\end{figure*}

The amplitude and phase of the first harmonic of the modulation are a measure of
the scalar product of drift velocity $\vec u_{\text{d}}$ and oscillation
velocity $\vec u_{\text{osc}}$ ~(\cref{eq:eta1}). There are two reasons for
$\vec u_{\text{d}}\cdot \vec u_{\text{osc}}$ to be nonzero. The first one is
the drift velocity from the ambipolar diffusion in the $xy$-plane and the
oscillation velocity in the same plane (due to the inductive coupling). However,
this should be zero only in the center and become nonzero everywhere else,
because the $u_{\text{amp}}$ is only zero in the center and~\cref{fig:eta2_amp}
shows, that the oscillation velocity is nonzero everywhere. The second one is
the drift velocity from ambipolar diffusion in the $z$-plane and the oscillation
velocity in that plane due to capacitive coupling. A first harmonic caused by
this may have any form. The amplitude and phase of the first harmonic of the
modulation, again at \SI{15}{\pascal} and \SI{700}{\watt} are shown
in~\cref{fig:eta1}. There is a vertical line, where the amplitude is zero and
the phase has a jump of about \SI{180}{\degree}. This line passes through the
center of the discharge. Therefore, it is more  likely that the first harmonic
here corresponds to the strength of the capacitive coupling.

Such a spatial dependence of the capacitive coupling in the INCA discharge
corresponds to the design of the discharge. An additional capacitor between
antenna and ground was added to decrease the capacitive
coupling~\cite{Ahr2018a}. This adds a virtual ground point in the middle of the
three arms of the antenna. At this ground point the RF potential should be
approximately zero, so there should be no capacitive coupling. Also there should
be a phase jump of \SI{180}{\degree}, because the RF potential changes its sign
at that position. This corresponds extremely well with the results from the
RFMOS measurement.

\begin{figure}[htb]
  \centering
  \includegraphics[width=\columnwidth]{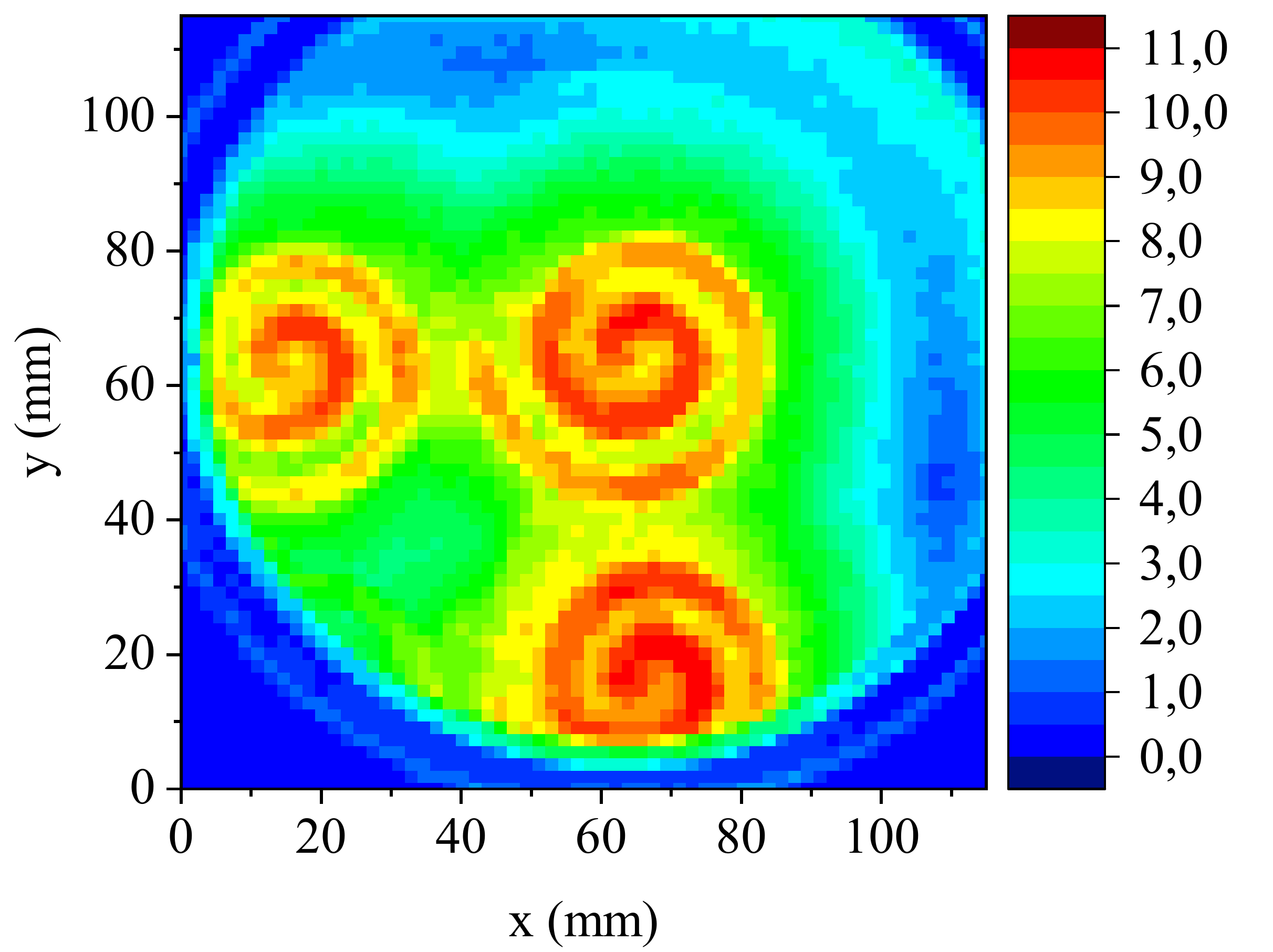}
  \caption{Amplitude of the first harmonic of the temporal modulation of the
    H$_\alpha$ line in a hydrogen plasma at \SI{15}{\pascal} and \SI{700}{\watt}
  measured through the upper front window. The color plot represents the
  amplitude in percent.}%
  \label{fig:eta1_upper}
\end{figure}

To check that the vanishing capacitive coupling only occurs in the (horizontal)
center of the discharge another measurement at \SI{15}{\pascal} was performed
at the upper front window. The amplitude of the first harmonic at that position
is shown in~\cref{fig:eta1_upper}. Note that the scale of these two figures is
not the same. The maximum of the amplitude in the central region is about
\SI{5.4}{\percent}, whereas the one in the corner region is significantly
higher, up to about \SI{11}{\percent}. The structure of the first harmonic
measured in the corner of the discharge supports the previous considerations.
Here the amplitude of the first harmonic is much stronger, no line of zero
capacitive coupling is visible and the coils form the maximum. All this
indicates that the capacitive coupling of the INCA discharge is consistent with
its design.

This also points out another property of the INCA discharge. Due to the virtual
ground point in the horizontal center of the antenna, the highest voltage does
not occur in the center of the discharge but at the left and right side of it,
which increases the capacitive coupling there. Also the electron density is low
there, because of the diffusion profile~\cite{Ahr2018}, which again enhances the
capacitive coupling. This is another possible explanation for the surprisingly
high plasma potential, which was attributed to an overpopulation of highly
energetic electrons~\cite{Ahr2018}. To confirm one or the other theory it would be
necessary to measure the density of electrons overcoming the sheath potential at
least quantitatively.

\section{Conclusion\label{sec:conclusion}}

Two different diagnostics were used to measure the electric field structure of
the antenna array of the INCA discharge. The more straightforward one, the B-dot probe, provides only \textit{ex situ} measurements while the other one, the radio frequency modulation spectroscopy, provides directly \textit{in situ} results. The results from the two diagnostic methods agree very well with each other. A criterion for the consistency of the
B-dot measurements was developed and applied to the measurements.
The structure of the electric field is far from ideal, hinting to the need of improved coil design that could lead to an improvement of the heating efficiency. Nevertheless, the measured
electric field structure is well in line with earlier theoretical
estimations~\cite{Ahr2018a}. A possible design for more ideal coils could be
one, that is still spiral but where the inner diameter is larger. This would
require the coils to be slightly narrower, but the stray current in radial
direction would also be smaller.

The strength of the capacitive coupling in the INCA discharge was also
determined with spatial resolution. It is confirmed that the designated virtual ground
point works as intended. This also leads to another explanation for the
surprisingly high plasma potential, which could be tested by measuring the
density of electrons overcoming the sheath potential.

\bibliography{inca-e}

\end{document}